\documentclass[twocolumn, prx, superscriptaddress]{revtex4-1}

\usepackage{amssymb}
\usepackage{amsmath}
\usepackage{epsfig}
\usepackage{color}
\usepackage{graphics, graphicx}
\usepackage{bbold}
\usepackage{psfrag}
\usepackage{mathcomp}
\usepackage{subfigure}
\usepackage{verbatim}
\usepackage{braket}
\usepackage{color}
\usepackage[colorlinks,citecolor=blue]{hyperref}
\def\cp#1{\mathbf{#1}}

\begin{document}
\title{Strongly Repulsive 1D Gases at Higher Branches: \\
Spin-Charge Correlation and Coupled Spin-Chain Model }
\author{Yu Chen}
\affiliation{Beijing National Laboratory for Condensed Matter Physics, Institute of Physics, Chinese Academy of Sciences, Beijing 100190, China}
\affiliation{School of Physical Sciences, University of Chinese Academy of Sciences, Beijing 100049, China}
\author{Xiaoling Cui}
\email{xlcui@iphy.ac.cn}
\affiliation{Beijing National Laboratory for Condensed Matter Physics, Institute of Physics, Chinese Academy of Sciences, Beijing 100190, China}
\date{\today}

\begin{abstract}

We investigate the higher repulsive branches of one-dimensional (1D) bosonic and fermionic quantum gases beyond the super-Tonks-Girardeau regime, utilizing the Bethe-Ansatz method and exact diagonalization of small trapped clusters. In contrast to the well-studied lowest branches that are characterized by spin-charge separation, we demonstrate the emergence of strong spin-charge correlation in all higher branches with hard-core interactions. This manifests in distinct quasi-momentum distributions and energy spectra for bosons and spin-1/2 fermions, despite their fermionization. Furthermore, trapped fermions in higher branches exhibit novel spin textures, intricately linked to  charge excitations, necessitating  a coupled multi-chain description beyond single effective spin-chain models. Our findings unveil a rich interplay between spin and charge degrees of freedom in highly excited 1D systems, opening avenues for exploring novel quantum phenomena beyond the conventional paradigm of low-lying states.

\end{abstract}

\maketitle

\section{Introduction}

One-dimensional (1D) ultracold atoms have offered an ideal platform for quantum simulation of strongly correlated phenomena, exhibiting markedly distinct behaviors compared to higher-dimensional systems\cite{1D_review1,1D_review2,1D_review3}. Notably, they can support a stable repulsive branch across a scattering resonance, effectively realizing the hard-core limit. In this regime, the system continuously evolves from the Tonks-Girardeau (TG) state (with coupling $g\rightarrow+\infty$)\cite{Girardeau1} to the super-Tonks-Girardeau (sTG) regime ($g\rightarrow-\infty$)\cite{QMC,BA,Chen, Chen_Guan,Girardeau2} without decaying to lower-lying attractive branches. This continuous evolution has been experimentally observed in atomic gases of identical bosons\cite{sTG_Science2009} and spin-1/2 fermions\cite{sTG_fermion_expt1, sTG_fermion_expt2}. In the hard-core limit ($g\rightarrow \infty$), or the intersection between TG and sTG regimes, these systems are fermionized with a frozen charge distribution, while any spin degrees of freedom remain free. Their wavefunctions can be factorized as
\begin{equation}
\Psi(\{x_i\},\{\xi_i\})=\psi_c(\{x_i\})\psi_s(\{x_i\},\{\xi_i\}). \label{sc_separation}
\end{equation} 
Here $x_i$ ($\xi_i$) represents the coordinate (spin) of the $i$-th atom; $\psi_c$ is the charge wavefunction given by the Slater determinant of free fermions, which is fully anti-symmetric under the permutation of any two particles;  $\psi_s$ describes the spin order on the 1D line, which involves sign functions of coordinates (see Eq.(\ref{spin_order})) in order to ensure the full symmetry of $\Psi$ upon simultaneously exchanging the spin and coordinate of any two particles. The factorization in Eq.(\ref{sc_separation}) explicitly demonstrates the spin-charge separation, as recently explored in spin-1/2 fermions\cite{Hulet}, in this specific hard-core regime. This separation leads to a large spin degeneracy since the energy is determined solely by the charge component. Consequently, the spin part can be conveniently manipulated  by small perturbations, resulting in rich quantum magnetism described by effective spin-chain models\cite{Zinner, Santos, Pu, Cui2, Parish, Zinner3, Parish2, BF_1,BF_2,suN_1,suN_2,suN_3}. For instance, spin-1/2 fermions follow an antiferromagnetic spin chain along the TG-sTG crossover\cite{sTG_fermion_expt2, Zinner, Santos, Pu, Cui2}, and various magnetic orders can be engineered around the hard-core limit by applying external perturbations\cite{soc_1,soc_2,pwave_1,pwave_2,pwave_3,local_mag_1,local_mag_2}.

Recently, an experimental breakthrough achieved access to higher repulsive branches of 1D atomic gases far beyond the TG/sTG regime\cite{sTG_dipole_expt}. By introducing a weak dipolar repulsion, the 1D gas exhibited remarkable stability over multiple interaction cycles across resonances. During this process, the system adiabatically entered higher repulsive branches with continuously increasing energy. This experiment overcomes the long-standing issue of severe atom loss in the sTG regime with negative coupling strength\cite{sTG_Science2009,sTG_fermion_expt1, sTG_fermion_expt2}. Remarkably, it was observed that the weak dipolar interaction significantly changes the stability of repulsive branch, but hardly affects its spectrum. The underlying mechanism of these observations has recently been revealed as the consequence of distinct energy responses, between the repulsive gas and its decay channels, to a weak dipolar force\cite{sTG_dipole_theory}. Importantly, the experimental achievement in \cite{sTG_dipole_expt} provides an unprecedented opportunity to explore novel 1D physics in highly excited states, which possess even stronger correlations than the TG/sTG states in the lowest branch. A key theoretical question is how to describe these strongly repulsive higher-branch states. Specifically, in the hard-core limit, do they still obey spin-charge separation as Eq.(\ref{sc_separation}), and what are the general rules governing charge and spin distributions in these highly excited states?

In this work, we address these fundamental questions by exactly solving the higher repulsive branches of 1D bosons and spin-1/2 fermions using Bethe ansatz method and exact diagonalization of small trapped clusters. Focusing on the hard-core limits of these higher branches, we find that they no longer feature spin-charge separation as Eq.(\ref{sc_separation}) but instead show strong spin-charge correlation. This correlation leads to new rules of the energetics and spin structures beyond our existing knowledge of hard-core systems. Take the homogeneous case for example, the higher branches of identical bosons and spin-1/2 fermions exhibit distinct quasi-momentum distributions and energy spectra, directly demonstrating the non-separable nature of spin and charge degrees of freedom in the latter.
To clearly visualize the spin structure of higher-branch fermions, we perform exact calculations on trapped ($1+N$) clusters consisting of one $\downarrow$ and $N$ $\uparrow$ fermions. These calculations reveal novel spin textures in the higher branches, qualitatively different from those observed in the lowest branch. The distinct textures are closely linked to charge excitations, providing further evidence of spin-charge correlation in trapped systems. Notably, due to  charge excitations, these higher branches cannot be described by a single effective spin chain but call for a coupled multi-chain treatment. From exact cluster solutions, we have conjectured a general principle governing the energetics and spin textures of hard-core ($1+N$) systems for arbitrary $N$. These results highlight an intriguing interplay between spin and charge degrees of freedom in the higher repulsive branches of 1D systems, suggesting the possibility of realizing exotic magnetic orders and correlated phases that are inaccessible in their low-lying counterparts.

The remainder of this paper is organized as follows. Section \ref{sec_model} presents the theoretical model. Section \ref{sec_homo} details the exact solutions for homogeneous systems using the Bethe ansatz, focusing on the differences between identical bosons and spin-1/2 fermions in their higher branches. Section \ref{sec_trap} examines $(1+N)$ fermions in a harmonic trap, highlighting the distinct spin structures of higher branches and introducing a coupled multi-chain model near the hard-core limit. Finally, Section \ref{sec_summary} provides the summary and outlook of our work.

\section{Model} \label{sec_model}

We consider the following Hamiltonian for 1D systems with contact interaction ($\hbar=1$):
\begin{equation}
H=\sum_i \left( -\frac{1}{2m} \frac{\partial^2}{\partial x_i^2} + \frac{1}{2} m\omega^2 x_i^2 \right)+g\sum_{\langle i,j\rangle} \delta(x_{i}-x_j); \label{H} 
\end{equation}
here $x_i$ is the 1D coordinate; $\omega$ is the frequency of harmonic trap; $g=-2/(m a)$ is the  coupling strength with 1D scattering length $a$. Note that for spin-1/2 fermions, the contact interaction only exists between different spins ($\uparrow$ and $\downarrow$) due to the symmetry requirement. Although the dipolar interaction is important for the stabilization of repulsive branch, we have not explicitly included it in our calculation since it hardly affects the spectrum of this branch\cite{sTG_dipole_expt,sTG_dipole_theory}.

In this work, we will study both homogeneous ($\omega=0$) and inhomogeneous ($\omega>0$) systems. For the homogeneous case, we employ the Bethe-ansatz  to analyze higher repulsive branches of identical bosons and spin-1/2 fermions with equal particle number. We have taken the periodic boundary condition and used the system size $L$ as the length unit.  For the inhomogeneous case,  we consider the harmonically trapped $(1+N)$ system consisting of one atom and $N$ identical fermions, using exact diagonalizations for small clusters with $N=2,3$ and further extrapolating to large systems with arbitrary $N$. We are particularly interested in the energetics and spin structure of higher repulsive branches in the hard-core limit. In the trapped case we take $l=\sqrt{2/(m\omega)}$  as the length unit.

\bigskip

\section{Higher branches of homogeneous systems} \label{sec_homo}

In this section, we study the repulsive branches of 1D bosons and spin-1/2 fermions in homogeneous systems with periodic boundary condition, which can be exactly solved using  Bethe-ansatz method\cite{BA, Chen_Guan, Chen}. For direct comparison, we consider the two systems with the same total number ($N$), and the fermions are spin-balanced with zero polarization ($N_{\uparrow}=N_{\downarrow}=N/2$).  

\begin{widetext}

\begin{figure}[t]
\includegraphics[width=17cm]{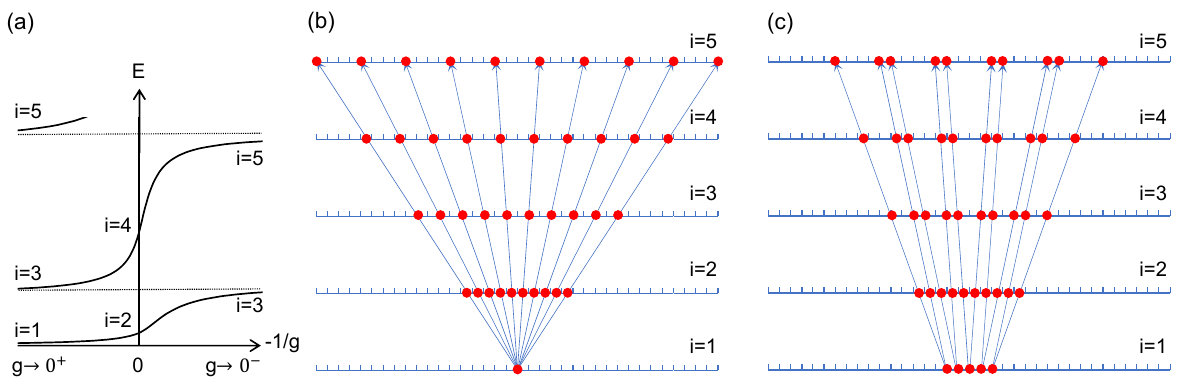}
\caption{(Color online) Adiabatic evolution of energies and quasi-momentum distributions for the homogeneous systems of identical bosons and spin-1/2 fermions in 1D. (a) is the schematics of energetic trajectory as changing $-1/g$, with indices $i=1,2,...$ marking the  repulsive branch at  special interaction limits ($g=0,\infty$) during  adiabatic evolution. (b) and (c) show quasi-momentum distributions of identical bosons and spin-1/2 fermions for  $i$ from $1$ to $5$. Both systems have the same particle number $N=10$, and the fermions are with zero polarization ($N_{\uparrow}=N_{\downarrow}=5$). The momentum is in unit of $2\pi/L$, with $L$ the system length. 
} \label{fig_BA}
\end{figure}

\end{widetext}

Fig.\ref{fig_BA}(a) shows the energetic trajectory of repulsive branch following an adiabatic evolution starting from the non-interacting regime ($g\rightarrow 0^+$). During the adiabatic process, the quasi-momentum distribution of repulsive branch evolves continuously, which is also the criterion for identifying the energy trajectory in Fig.\ref{fig_BA}(a).  For clarity, we focus on two interaction strength limits: non-interacting ($g=0^{\pm}$) and hard-core ($g=\pm \infty$), as marked by indices $i=1,2,...$ in Fig.\ref{fig_BA}(a) along the adiabatic trajectory. In the experiment of \cite{sTG_dipole_expt}, the system was driven through two full interaction circles, i.e., from $i=1$ to $i=5$.  Fig.\ref{fig_BA}(b1,b2) show the corresponding quasi-momentum distributions $\{k_j,\ j=1,...N\}$ for the repulsive branch at each index $i$, for both identical bosons and spin-1/2 fermions with the same total number $N=10$. We can see that 
the two systems  exhibit distinct evolutions of $\{k_j\}$ as $i$ increases.  

For identical bosons, it has been shown that  neighboring quasi-momenta ($k_j$ and $k_{j+1}$) are always equally spaced, and the spacing increases linearly with $i$\cite{Cheon}. Specifically, at a given $i$, the quasi-momenta are given by $k_j=(2\pi/L)(i-1)(j-(N+1)/2)$.  
For a thermodynamic system with fixed density $n=N/L$ (where $N,L\rightarrow \infty$), the energy per particle for repulsive bosons at index $i$ is
\begin{equation}
\epsilon_B^{(i)}=(i-1)^2\epsilon_F,\ \ \ \ {\rm with}\ \ \ \epsilon_F=\frac{\pi^2n^2}{6m}. \label{e_b}
\end{equation}
For spin-1/2 fermions, however, $\{k_j\}$ are equally spaced only in the  hard-core limit of the lowest branch (at index $i=2$). For all higher branches, they are no longer equally spaced but always emerge as pairs. In the limit of $N\rightarrow \infty$, there are $N/2$ pairs of $\{k_j,k_{j+1}=k_j+2\pi/L\}$, and the inter-pair spacing increases linearly with $i$. This yields the energy per particle for fermions at index $i$: 
\begin{equation}
\epsilon_F^{(i)}=\frac{i^2}{4}\epsilon_F. \label{e_f}
\end{equation}

Comparing Eq.(\ref{e_b}) with  Eq.(\ref{e_f}), we can see that identical bosons and spin-1/2 fermions generally have different energies and quasi-momentum distributions.  In this sense, the lowest branch in hard-core limit ($i=2$) appears as a very special case where the two systems have the same $\{k_j\}$ and the same energy due to spin-charge separation (see Eq.\ref{sc_separation}). This is also known as Bose-Fermi mapping\cite{Girardeau1} or Fermi-Fermi mapping\cite{Guan, Girardeau2, Blume}, which tells that the hard-core bosons and spin-1/2 fermions can all be mapped to identical fermions with the same charge distribution, resulting in equivalent  energy and quasi-momentum distribution for the same particle number. However,  such equivalence breaks down for the hard-core limit of higher branches (with $i=4,6,..$), where the fermionized bosons have a much higher energy than the fermionized spin-1/2 fermions, see Eqs.(\ref{e_b}, \ref{e_f}). Therefore, the higher branches can distinguish well between bosonic and fermionic systems in their charge distributions even  both of them are fermionized.   In particular, the pairwise $\{k_j,k_{j+1}\}$ distributions of higher-branch fermions, as shown in Fig.\ref{fig_BA}, suggest an intricate correlation between spin and charge degrees of freedom. Unfortunately, it is extremely hard to figure out an analytical form of spin-charge wavefunction  from  Bethe-ansatz solutions of higher-branch fermions. In the next section, we will turn to harmonically trapped system that is more relevant to realistic experiment, where the spin-charge correlation can also be viewed much more transparently.

\section{Higher branches of trapped systems} \label{sec_trap}

To clearly  visualize the spin structures of higher repulsive branches, in this section we exactly solve the harmonically trapped small clusters, from which we will extract general rules for large systems. Due to the limitation on numerical computation, here we consider ($1+N$) clusters that consist of one atom and $N$ identical fermions, which requires much less numerics than ($N+N$) clusters with the same $N(>1)$. Such $(1+N)$ systems are also known as Fermi polarons as studied extensively in literature (see recent reviews\cite{polaron_review1, polaron_review2}).  The exact solutions of   $(1+N)$ clusters with $N=2, 3$ have also been studied before\cite{Zinner, Santos, Pu, Parish, Zinner2,Blume, Conduit,sTG_dipole_theory}. Different from all these studies,  here we will focus on the higher repulsive branches  following the adiabatic trajectory shown in Fig.\ref{fig_BA}. In appendix \ref{appendix_cluster}, we have presented the formula for exactly solving ($1+2$) and ($1+3$) problems. 


Before proceeding, let's first recall the effective spin-chain model for the lowest branch of spin-1/2 fermions near hard-core limit\cite{Zinner, Santos, Pu, Cui2, Parish}:
\begin{equation}
H_{\rm sc}=\sum_j \frac{J_j}{g} {\cp s}_j\cdot {\cp s}_{j+1}, \label{single_chain}
\end{equation}
where $j$ indexes the  spin order in real space, ${\cp s}_j$ is the Pauli operator for atom at the $j$-th order, and  $J_j$ is the exchange coupling strength associated with charge distributions. For an $n$-body system, a spin-ordered state, $\ket{\xi_1\xi_2 \cdots \xi_n}$,  describes a sequence of spins $\xi_{1}$, $\xi_{2},\cdots,\xi_{n}$ placed in order on the 1D chain. Explicitly, its wavefunction reads
\begin{eqnarray}
&&\langle x_1,\cdots,x_n; \mu_1,\cdots,\mu_n | \xi_1\xi_2 \cdots \xi_n\rangle \nonumber\\
&=&\sum_{P}
 \theta(x_{P_1}< x_{P2}<\cdots<x_{P_n})
\prod_i
 \delta_{\xi_i,\mu_{P_i}},  \label{spin_order}
\end{eqnarray}
where $P$ is a permutation of $(1,2, \cdots, n)$, and $\theta(x_{P_1}< x_{P2}<\cdots<x_{P_n})$ is non-zero $(=1)$ only for $x_{P_1}< x_{P2}<\cdots<x_{P_n}$. The spin-ordered state (\ref{spin_order}) exactly comprises the spin part of the wavefunction in Eq.(\ref{sc_separation}). 

To facilitate later discussions on small clusters, we now write down the relevant eigen-states of Eq.(\ref{single_chain}). For ($1+2$) system, there are two orthogonal eigen-states with total spin $S=1/2$:
\begin{eqnarray}
\ket{1}&=&\frac{1}{\sqrt{6}} \left(\ket{\downarrow\uparrow\uparrow}-2\ket{\uparrow\downarrow\uparrow}+\ket{\uparrow\uparrow\downarrow}\right);\nonumber\\
\ket{2}&=&\frac{1}{\sqrt{2}} \left(\ket{\downarrow\uparrow\uparrow}-\ket{\uparrow\uparrow\downarrow}\right).\label{3_eigen}
\end{eqnarray} 
For ($1+3$) system, there are three eigen-states with $S=1$:
\begin{eqnarray}
\ket{1}&=& C_-\ket{\downarrow\uparrow\uparrow\uparrow}-C_+\ket{\uparrow\downarrow\uparrow\uparrow}+C_+\ket{\uparrow\uparrow\downarrow\uparrow}-C_-\ket{\uparrow\uparrow\uparrow\downarrow};\nonumber\\
\ket{2} &=& \frac{1}{2}\left(\ket{\downarrow\uparrow\uparrow\uparrow}-\ket{\uparrow\downarrow\uparrow\uparrow}-\ket{\uparrow\uparrow\downarrow\uparrow}+\ket{\uparrow\uparrow\uparrow\downarrow}\right);\nonumber\\
\ket{3}&=&C_+\ket{\downarrow\uparrow\uparrow\uparrow}+C_-\ket{\uparrow\downarrow\uparrow\uparrow}-C_-\ket{\uparrow\uparrow\downarrow\uparrow}-C_+\ket{\uparrow\uparrow\uparrow\downarrow}, \label{4_eigen}
\end{eqnarray}
where $C_\pm=\frac{1}{2}\sqrt{1\pm\frac{J_2}{\sqrt{J_1^2+J_2^2}}}$. Note that the spin states in Eqs.(\ref{3_eigen},\ref{4_eigen}) only apply to systems with open boundaries, excluding the case of  homogeneous system with periodic boundary condition. 
 Previous studies of  $(1+2)$ and $(1+3)$ clusters in a harmonic trap showed that their lowest branches in hard-core limit follow  state $\ket{1}$ in Eqs.(\ref{3_eigen},\ref{4_eigen})\cite{Zinner, Santos}. 

In the following, we will first present exact results of trapped ($1+2$) and ($1+3$) clusters,  highlighting the distinct spin structures of higher branches compared to the lowest one. Further, we construct a coupled spin-chain model for the higher branches in hard-core limit, which features a strong correlation between spin and charge degrees of freedom. Based on these results,  we will finally extract a general rule for the energetics and spin textures of  $(1+N)$ system with arbitrary $N$.

\subsection{Exact results of ($1+N$) clusters}

In Fig.\ref{fig_3body}(a) and Fig.\ref{fig_4body}(a), we plot out the energy spectra  of ($1+2$) and ($1+3$) systems from exact diagonalizations, where the red curves show the trajectories of repulsive branches under adiabatic evolution from the non-interacting regime. Note that the adiabatic repulsive branch can have level crossing with other branches at the locations of energy degeneracy, such as in the hard-core or non-interacting limits, and in these cases we have uniquely identified the adiabatic branch by the continuity of energy slope (i.e., the derivative of $E$ with respect to $1/g$ or $g$). Meanwhile, at attractive couplings ($g<0$), the repulsive branch can have avoided level crossings with branches of excited bound states. In these cases we have identified the adiabatic trajectory by comparing the wavefunctions before and after the avoided crossings, as followed from  \cite{sTG_dipole_theory} (see Fig.2 therein).

On the red curves in Fig.\ref{fig_3body}(a) and Fig.\ref{fig_4body}(a), the hard-core limits of these repulsive branches are marked by  'A,B,C' from low to high branches. In Fig.\ref{fig_3body}(A,B,C) and Fig.\ref{fig_4body}(A,B,C), we further show the corresponding wavefunctions $\Psi(x_2-x_1,x_3-x_1)$, with $x_1$ and $x_{2,3}$ respectively the coordinates of (single) $\downarrow$ and (multiple) $\uparrow$ fermions. Note that in plotting $\Psi(x_2-x_1,x_3-x_1)$ for $(1+3)$ system, we have integrated over the relative motion between the remaining $\uparrow$ fermion ($x_4$) and the $\downarrow$ atom ($x_1$). In this way, $\Psi(x_2-x_1,x_3-x_1)$  directly shows the $\uparrow$-$\downarrow$ and $\uparrow$-$\uparrow$ correlation patterns and reflects the spin texture of the system. 

For the lowest branch near  hard-core limit (A), its wavefunction exhibits spin-charge separation as in Eq.(\ref{sc_separation}). In this case, the charge distribution is frozen at the ground state of identical fermions, and the spin part  can be effectively described by a single spin chain in Eq.(\ref{single_chain}). Indeed, we have confirmed that for both ($1+2$) and ($1+3$) systems, their lowest branches at (A) follow state $\ket{1}$ in Eqs.(\ref{3_eigen},\ref{4_eigen}) from the spin-chain model. This state has the largest energy slope across hard-core limit, or equivalently, the largest 1D contact\cite{contact_1D}
\begin{equation}
C=\frac{\partial E}{\partial (-1/g)}.
\end{equation} 
The fact that $\ket{1}$ has the largest $C$ is because in $\ket{1}$, the $\downarrow$ atom dominantly stays at the trap center and thus experiences the largest exchange coupling with neighboring $\uparrow$ fermions. Accordingly, its  wavefunction $\Psi(x_2-x_1,x_3-x_1)$, as shown in Fig.\ref{fig_3body}(A) and Fig.\ref{fig_4body}(A), has the largest weight when $(x_2-x_1)(x_3-x_1)<0$, i.e., when  $\downarrow$ ($x_1$) stays in-between two $\uparrow$ fermions ($x_2$ and $x_3$). 

Remarkably, the higher branches (B) and (C) exhibit distinct spin textures as compared to the lowest branch. As shown in Fig.\ref{fig_3body}(B,C) and Fig.\ref{fig_4body}(B,C), for these higher branches  $\Psi$ has the largest weight when $(x_2-x_1)(x_3-x_1)>0$, i.e., $\downarrow$ tends to reside at the left or right sides of all $\uparrow$ fermions. These states appear to correspond to $\ket{2}$ in Eq.(\ref{3_eigen}) and $\ket{3}$ in Eq.(\ref{4_eigen}), which have the smallest contact among all relevant spin states with the same $S(=\frac{N-1}{2})$. 

As we will discuss in the next subsection, the dramatic change of spin textures for all higher branches are deeply rooted in a strong correlation between spin and charge degrees of freedom, where the single spin-chain model in Eq.(\ref{single_chain}) becomes invalid for their description. 

\begin{widetext}

\begin{figure}[h]
\includegraphics[width=17cm]{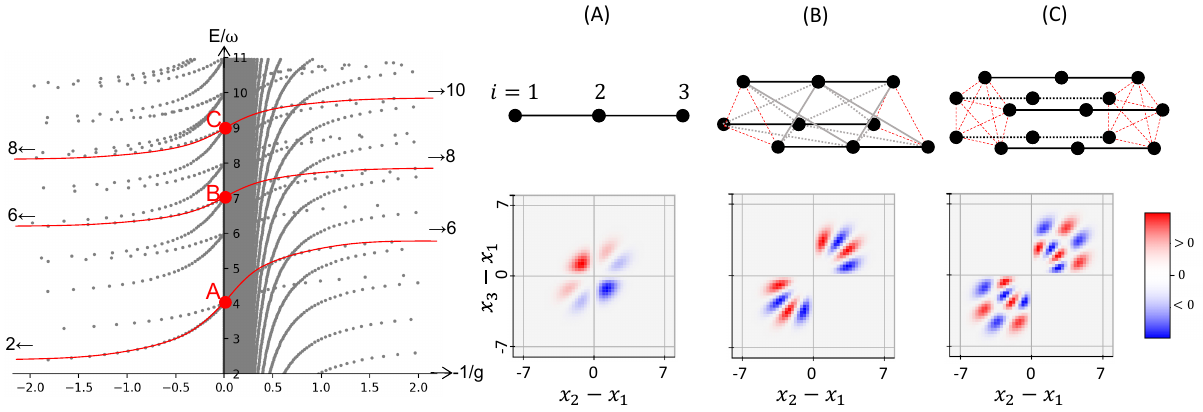}
\caption{(Color online) Exact solutions of harmonically trapped $(1+2)$ system in 1D that consists of one $\downarrow$ and two $\uparrow$ fermions. Left side is the Energy spectrum, where the red curves denote the adiabatic trajectory of repulsive system as changing interaction strength, and 'A',  'B', 'C' mark the hard-core limits of the lowest and higher branches. In the right side, (A,B,C) show the corresponding effective spin chains and wavefunctions $\Psi(x_2-x_1,\ x_3-x_1)$, where $x_1$ and $x_{2,3}$ are respectively the coordinates of $\downarrow$ and $\uparrow$ fermions. The length unit is $l=\sqrt{2/(m\omega)}$. For the lowest branch (A), the system is described by a single spin chain as Eq.(\ref{single_chain}) and the spin state follows $|1\rangle$ in Eq.(\ref{3_eigen}). For higher branches (B) and (C), they are described by coupled spin chains, forming the geometries of triangular prism and pentaprism respectively. The spin states of higher branches follow $|2\rangle$ in Eq.(\ref{3_eigen}).  } \label{fig_3body}
\end{figure}

\begin{figure}[t]
\includegraphics[width=17cm]{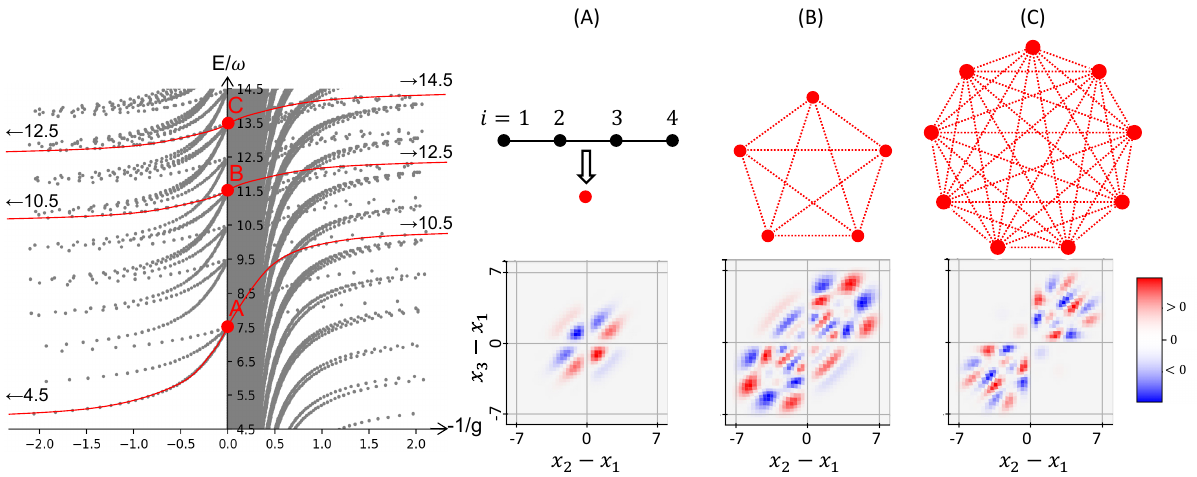}
\caption{(Color online) Same as Fig.\ref{fig_3body} except for $(1+3)$ system. In plotting $\Psi(x_2-x_1,\ x_3-x_1)$ in (A,B,C), we have integrated over the relative motion between the remaining fermion ($x_4$) and the impurity ($x_1$). The lowest branch (A) follows spin state $\ket{1}$ in Eq.(\ref{4_eigen}) as described by a single spin chain. Upon spin reduction, each chain can be further simplified as a (red) point.  For higher branches (B) and (C), the system can be effectively described by a coupled point model in Eq.(\ref{coupled_point}), forming the linked pentagon and nonagon respectively. The spin structures of these higher branches essentially follow state  $\ket{3}$ in Eq.(\ref{4_eigen}). } \label{fig_4body}
\end{figure}

\end{widetext}

\subsection{Coupled spin-chain model for higher branches}

Different from the lowest branch (A), the higher branches (B) and (C) are no longer associated with a unique charge distribution due to finite excitation energies. Consequently, they cannot be  described by a single spin chain model in Eq.(\ref{single_chain}). For these higher branches, there are  multiple charge configurations and each charge configuration is associated with a single spin chain. The exchange of two neighboring spins can occur within each chain and between different chains. The associated effective model can be derived straightforwardly by treating the interaction as zero-th order Hamiltonian and the kinetic term as perturbations\cite{Pu}, which leads to a  coupled multi-chain model as:
\begin{equation}
H_{\rm cc}= \sum_{j=1}^N \sum_{\langle\alpha,\beta\rangle} \frac{J^{(\alpha,\beta)}_j}{g} {\cp s}^{(\alpha)}_j\cdot {\cp s}^{(\beta)}_{j+1}, \label{coupled_chain}
\end{equation} 
where $\alpha,\beta$ are the chain (or charge) indices, and the exchange coupling $J^{(\alpha,\beta)}_j$ follows 
\begin{eqnarray}
J^{(\alpha,\beta)}_j&=&\frac{(N+1)!}{2m^2} \int d{\bf x}  \frac{\partial \psi_c^{(\alpha)*}}{\partial x_j} \frac{\partial \psi_c^{(\beta)}}{\partial x_j} \nonumber\\
&&\ \ \theta(x_1<\cdots<x_j=x_{j+1}<\cdots<x_{N+1}),    \label{J_coupled}
\end{eqnarray}
with $\psi_c^{(\alpha)}$ the charge wavefunction of according chain $\alpha$. For the single chain case ($\alpha=\beta=1$), Eq.(\ref{coupled_chain}) directly reduces to $H_{\rm sc}$ in Eq.(\ref{single_chain}).

Take the ($1+2$) system for example, the higher branch (B) has three charge configurations $\{n\}\equiv (n_1,n_2,n_3)=(0,1,5),\ (0,2,4),\ (1,2,3)$, and therefore it  corresponds to three coupled chains forming the geometry of triangular prism, see Fig.\ref{fig_3body}(B). For (C), it has five charge configurations $\{n\}=(0,1,7),\ (0,2,6),\ (0,3,5), \ (1,2,5),\ (1,3,4)$, and therefore it is associated with five couple chains forming a pentaprism, see Fig.\ref{fig_3body}(C). Apparently, higher branches have  more charge configurations. For a general $(1+N)$ system at a given higher branch with $n_c$ charge configurations, the coupled chain model (Eq.\ref{coupled_chain}) yields  $n_c N$ eigen-states. The key question is which eigen-state is to be selected by the adiabatic trajectory of repulsive branch. To answer this question, a general selection rule has to be identified. Such selection rule, as we will discuss below, is determined by the spin-charge correlation.

The spin-charge correlation of higher branches originates from the conservation of total parity of the whole system. Let's denote $P_c$ ($P_s$) as the charge (spin) parity, which is either $1$ or $-1$ depending on whether the according charge (spin) wavefunction changes signs under mirror reflection (i.e., all coordinates $x_i\rightarrow-x_i$). For instance, spin states $\ket{1}$ and $\ket{2}$ in Eq.(\ref{3_eigen}) have $P_s=1$ and $-1$  respectively, and $\ket{1}$, $\ket{2}$, $\ket{3}$ in Eq.(\ref{4_eigen}) have $P_s=-1,\ 1, \ -1$ respectively.  The total parity is then given by 
\begin{equation}
P=P_cP_s. 
\end{equation}
During adiabatic evolution of the repulsive system, $P_c$ and $P_s$ may change individually but $P$ is conserved. Therefore, once $P$ is determined by the lowest branch, it remains the same for all higher branches, leading to a constraint between the spin and charge degrees of freedom. Take $(1+2)$ system for example, we can see that the lowest branch (A) (following $\ket{1}$) has $P_c=-1$, $P_s=1$ and total $P=-1$. For the higher branches (B) and (C), the charge excitation energies are respectively $\Delta E_{AB}=3\omega$ and $\Delta E_{AC}=5\omega$, and thus their charge parities  both switch to $P_c=1$. To maintain $P=-1$, the spin parity must change to $P_s=-1$, for which $\ket{2}$ is the only option. In this sense, the parity conservation build up a link between spin and charge and make them strongly correlated with each other in all higher branches. 

Similar analysis also applies to $(1+3)$ system. For the lowest branch we have $P_c=1$, $P_s=-1$, again giving $P=-1$. From the lowest (A) to higher (B,C) branches, the charge excitation energies are $4\omega$ and $6\omega$, and thus $P_c=1$ is unchanged. To keep $P=-1$, all higher branches must have $P_s=-1$, and two states in Eq.(\ref{4_eigen}), $\ket{1}$ and $\ket{3}$, satisfy this condition. On the other hand, from the energy spectrum in Fig.\ref{fig_4body}(a), we can see that all higher branches within an interaction circle (from $-1/g=-\infty$ to $+\infty$) experience the smallest energy shift ($=2\omega$). Therefore, one has to choose state $\ket{3}$ with the smallest contact across hard-core limit. In this way, the spin states changes dramatically  from $\ket{1}$ to $\ket{3}$, as the system evolves from the lowest to higher branches. This is why the spin textures of higher branches behave so differently from the lowest branch, see Fig.(\ref{fig_3body}) and Fig.(\ref{fig_4body}).

In above we have analyzed the spin change from $\ket{1}$ to other states within the single-chain framework, i.e., based on the eigen-states in Eqs.(\ref{3_eigen},\ref{4_eigen}) from the single spin chain model. While this single-chain picture provides physical insight, it is not quantitatively accurate for $(1+N)$ systems with $N>2$, given the presence of multi-chain configurations and the fact that each chain have its own eigen-states. In this case, we must consider the effect of inter-chain coupling. Based on all these analyses, we can simplify the multi-chain model in Eq.(\ref{coupled_chain}) as follows. First, we assume the spin of each chain is pinned at the state with conserved total parity and the smallest contact. Then, each chain can be reduced to a single point and the inter-chain coupling leads to the coupled point model:
\begin{equation}
H_{\rm cp}=  \sum_{\langle\alpha,\beta\rangle} \frac{C_{\alpha\beta}}{g} |\alpha\rangle\langle\beta|, \label{coupled_point}
\end{equation}
where $|\alpha\rangle,\ |\beta\rangle$ are point indices denoting various charge configurations with their according spin states, and $C_{\alpha\beta}$ is the strength of inter-point coupling involving the spin coefficients and exchange couplings. Take the $(1+3)$ system for example, the spin of each chain is pinned at state $\ket{3}$ in Eq.(\ref{4_eigen}), and therefore we have 
\begin{eqnarray}
|\alpha\rangle&=&C^{(\alpha)}_+\ket{\downarrow\uparrow\uparrow\uparrow}+C^{(\alpha)}_-\ket{\uparrow\downarrow\uparrow\uparrow}-C^{(\alpha)}_-\ket{\uparrow\uparrow\downarrow\uparrow}-C^{(\alpha)}_+\ket{\uparrow\uparrow\uparrow\downarrow}, \nonumber
\end{eqnarray}
with $C^{(\alpha)}_\pm$ the spin coefficients for chain $\alpha$ (see definition below Eq.\ref{4_eigen}). Given the coupled-chain model in Eq.(\ref{coupled_chain}), the coefficient $C_{\alpha\beta}$ in Eq.(\ref{coupled_point}) can be determined as
   \begin{eqnarray}
  C_{\alpha\beta}&=&4\left[C^{(\alpha)}_-C^{(\beta)}_+J_1^{\left(\alpha,\beta\right)}-C^{(\alpha)}_+(C^{(\beta)}_+-C^{(\beta)}_-)J_1^{\left(\alpha,\beta\right)}\right.
  \nonumber\\
  &&\left. -C^{(\alpha)}_-C^{(\beta)}_-(J_1^{\left(\alpha,\beta\right)}+2J_2^{\left(\alpha,\beta\right)})\right],
 \end{eqnarray}
For higher branches (B) and (C) of $(1+3)$ system, these points form  pentagon and nonagon respectively, as shown in Fig.\ref{fig_4body}(B,C). 
 
The coupled point model in Eq.(\ref{coupled_point}) can greatly simplify the numerical calculations of higher branches near hard-core limit, and has been found to reproduce their wavefunctions with high accuracies. For $(1+3)$ system, we have checked that this model can easily produce the higher-branch wavefunctions at (B) and (C) with accuracies $>99.8\%$ when compared to exact results. 

\subsection{Generalization to ($1+N$) system with large $N$}

Based on exact results of trapped $(1+N)$ clusters, here we consider systems with arbitrary $N$ and make  a conjecture on  their energetics and spin textures in hard-core limit.

\begin{figure}[t]
\includegraphics[width=9cm]{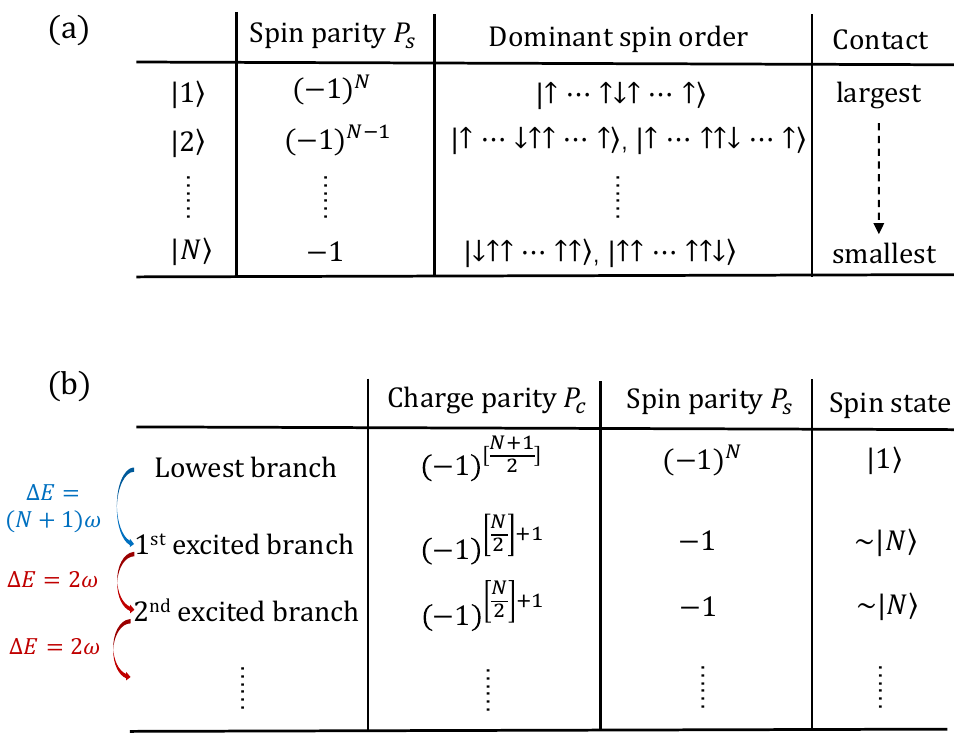}
\caption{General properties of harmonically trapped $(1+N)$ system with arbitrary $N$ in hard-core limit. (a) The spin parity, dominant spin order and contact of all relevant spin states for $(1+N)$ system with total spin $S=(N-1)/2$ and total parity $P=(-1)^{[\frac{N}{2}]}$. (2) The charge parity, spin parity and spin state of the lowest and higher branches of adiabatically evolving $(1+N)$ system in hard-core limit. The energy change between neighboring branches are also denoted ($\omega$ is the trap frequency). } \label{fig_table}
\end{figure}

First, let's look into the eigen-states of single spin-chain model (Eq.\ref{single_chain}). Here we focus on the states with total spin $S=(N-1)/2$, as followed by ($1+N$)  system under adiabatic evolution from non-interacting regime. 
With this total spin, there are $N$ orthogonal eigen-states of Eq.(\ref{single_chain}), labeled as $|i\rangle$ ($i=1,2..N$) in Fig.\ref{fig_table}(a). As $i$ increases, the according state changes gradually in both spin parity and dominant spin order. For $\ket{1}$, the impurity $\downarrow$ is predominantly at the trap center with spin parity $P_s=(-1)^N$, and its contact is the largest  due to the maximal spin exchange with surrounding $\uparrow$ fermions. As increasing $i$, $\downarrow$ gradually moves from the trap center to the edge, with decreasing contact and oscillating $P_s$. For  $\ket{N}$,   $\downarrow$ is predominantly at the trap edges with $P_s=-1$, and its contact is reduced to the smallest value.  For large $N$, one can imagine a substantial change of spin structure from $\ket{1}$ to $\ket{N}$. This is exactly the change occurring for the adiabatic repulsive system from the lowest to higher branches, ad discussed below. 

From the exact results of small $(1+N)$ clusters, we now conjecture on the general properties of hard-core systems with arbitrary $N$ under adiabatic evolution, as summarized in Fig.\ref{fig_table}(b). For the lowest branch, it is known that the charge is frozen at the ground state with  $\{n=0,1,...N\}$, whose parity is $P_c=(-1)^{[\frac{N+1}{2}]}$. Since the spin state of the lowest branch is $\ket{1}$ with parity $P_s=(-1)^N$, the total parity is then $P=(-1)^{[\frac{N}{2}]}$, which is conserved for all the higher branches. The conserved $P$ leads to strong spin-charge correlation in higher branches, namely, their spin textures are closely related to their charge excitations from the lowest branch. 

From the lowest to the first excited branch, the energy difference is $\Delta E=(1+N)\omega$, and therefore $P_c$ is changed to $(-1)^{[\frac{N}{2}]+1}$ in the first excited branch. Given the conservation of total parity $P$, this change in $P_c$ necessitates a change in spin parity to $P_s=-1$, directly manifesting the spin-charge correlation. Combined with the smallest energy shifts (or contacts) for all higher branches, we expect that the first excited branch corresponds to state   $\ket{N}$ for general $(1+N)$ systems, see Fig.\ref{fig_table}(a). For even higher branches, we expect the charge excitation energies between neighboring branches are always $\Delta E=2\omega$, the smallest energy shift ever achievable in a harmonic trap, and therefore  $P_s$ and $P_c$ are always unchanged from the first excited branch. This is to say, all higher branches maintain the same spin distribution  as the first excited branch. Note that due to the presence of multi-chain configuration, the specific form of $\ket{N}$ for different chains can be different. The actual spin states of higher branches can be deduced from the coupled point model in Eq.(\ref{coupled_point}), where each point refers to a particular $\ket{N}$ within each chain.   


\section{Summary and outlook} \label{sec_summary}

In summary, we have investigated the higher repulsive branches of 1D bosons and spin-1/2 fermions following the adiabatic trajectory starting from the non-interacting regime. Our focus has been on the hard-core limit of these excited branches, where we have uncovered a strong correlation between spin and charge degrees of freedom, in stark contrast to the spin-charge separation characteristic of the lowest branch. This fundamental difference distinguishes these highly excited states from previously studied 1D fermionized systems.
The emergent spin-charge correlation in higher branches manifests in distinct quasi-momentum distributions and energy spectra for fermionized bosons and fermions, as well as unique spin textures in trapped $(1+N)$ systems that are intricately linked to charge excitations. Notably, we have demonstrated that the spin textures of these higher branches in trapped systems cannot be described by a single spin-chain model due to the presence of multiple charge configurations. To address this, we have developed a coupled multi-chain model, which is further simplified to a computationally efficient coupled point model based on spin reduction. These effective models  provide a powerful tool for investigating the quantum magnetism of highly excited 1D systems in the strongly repulsive regime. Furthermore, through exact solutions of small clusters, we have established a general principle governing the energetics and spin structures of trapped higher-branch $(1+N)$ fermions for an arbitrary $N$.

Our findings are readily observable in degenerate quantum gases of dipolar mixtures, where a weak  dipolar repulsion is expected to stabilize the system during adiabatic evolution to higher branches\cite{sTG_dipole_theory}. Recent experimental advances have enabled the realization of Fermi-Fermi and Fermi-Bose dipolar mixtures in ultracold atomic systems with highly tunable interactions, achieved using different hyperfine states of the same atomic species\cite{expt_dipolar_1} or distinct atomic species\cite{expt_dipolar_2}. Quasi-one-dimensional (1D) confinement of these systems allows for further tuning of the 1D coupling strength via confinement-induced resonances\cite{CIR1,CIR2}. The characteristic spin textures predicted for the higher repulsive branches can be probed through tunneling\cite{sTG_fermion_expt2} or momentum-space\cite{expt_noise1,expt_noise2,expt_noise3,expt_noise4,expt_noise5,Jochim_expt2} measurements.

This work has revealed a robust spin-charge correlation in spin-1/2 fermions for two vastly different polarization regimes, i.e., the balanced systems in homogeneous settings ($P\equiv(N_{\uparrow}-N_{\downarrow})/(N_{\uparrow}+N_{\downarrow})\rightarrow 0$) and the highly polarized systems in traps ($P\rightarrow 1$). The presence of this correlation in such disparate conditions suggests its universality across a broad range of higher-branch 1D systems, including spin-1/2 fermions at arbitrary polarizations as well as  other mixtures, such as boson-boson, boson-fermion, and high-spin systems. The interplay between this spin-charge correlation, quantum statistics, and spin polarization offers a promising avenue for engineering exotic magnetic orders and correlated phenomena in these highly excited states, which could be hardly achieved in their low-lying counterparts. We hope our present work will stimulate further theoretical and experimental investigations into the rich physical consequences of this fundamental correlation.

Data that support the findings of this article are openly
available\cite{data}.

\bigskip

{\bf Acknowledgement.} We thank Xi-Wen Guan, Doerte Blume, Meera Parish and Jesper Levinsen for helpful discussions. This work is supported by the National Natural Science Foundation of China (92476104, 12134015).

\appendix



\section{Formula for exactly solving $(1+N)$ clusters in a harmonic trap} \label{appendix_cluster}

Here we present the formula for exactly solving $(1+N)$ problems in a harmonic trap, with $N=2,3$. We take $x_1$ as the coordinate of spin-$\downarrow$, and $x_2,x_3,...$ as the coordinates of spin-$\uparrow$ fermions.
We note that the same problems have been studied before\cite{Zinner, Santos, Pu, Parish, Zinner2,Blume, Conduit,sTG_dipole_theory}, mostly focusing on the lowest-branch physics. 

\subsection{Three-body ($N=2$)}

After decoupling the center-of-mass (CoM) motion from the problem, we  define the relative coordinates as
\begin{equation}
r=x_2-x_1,\ \ \ \ \rho=\frac{2}{\sqrt{3}}(x_3-\frac{x_1+x_2}{2}).
\end{equation} 
Similarly, we have another set of relative coordinates $\{r_+,\rho_+\}$ by exchanging $x_2\leftrightarrow x_3$ in $\{r,\rho\}$.  In the CoM frame, the three-body wave function can be expanded as 
\begin{equation}
	\Psi(r,\rho)=\sum_{mn} c_{mn}\phi_{m}(r)\phi_{n}(\rho),
	\label{function_expansion}
\end{equation}
with single-particle eigen-state
\begin{equation}
\phi_{n}(x)=\frac{1}{\pi^{\frac{1}{4}}\sqrt{2^nn!l}}e^{-\frac{x^2}{2l^2}}H_n(x/l),
\end{equation}
and eigen-energy $\epsilon_n=(n+1/2)\omega$. Here the trap length is defined as $l=\sqrt{2/(m\omega)}$. 

We rewrite the Hamiltonian (\ref{H}) as $H=H^{(0)}+U$, where $H^{(0)}$ and $U$  denote the non-interacting and interacting parts of $H$, respectively. Introducing an auxiliary function $f(r,\rho)\equiv U\Psi(r,\rho)$, and ensuring its anti-symmetry 
\begin{equation}
f(r,\rho)=-f(r_+,\rho_+), \label{f_symmetry}
\end{equation} 
we can write $f$-function as  
\begin{equation}
	\begin{split}
		f(r,\rho)=g\Big(&\sum_{mn} c_{mn}\phi_{m}(0)\phi_{n}(\rho)\delta(r) \\
		&- \sum_{mn} c_{mn}\phi_{m}(0)\phi_{n}(\rho_+)\delta(r_+)  \Big). 
	\end{split} \label{f}
\end{equation}
Further incorporating the Lippmann-Schwinger equation 
\begin{equation}
	\Psi=G_0 U \Psi, \label{LS}
\end{equation}
with $G_0=(E-H^{(0)})^{-1}$ the non-interacting Green's function, we obtain the self-consistent equation for $\{ c_{mn}\}$:
\begin{equation}
(E-\epsilon_m-\epsilon_n)c_{mn}=g\sum_{ij}c_{ij}\phi_i(0)\Big(\phi_m(0)\delta_{j,n} - A^{(1)}_{mn,j}\Big),  \label{fermi_eigen_equation1}
\end{equation}
where 
\begin{eqnarray*}
	A^{(1)}_{mn,j}&=&\int d\rho \phi_m(\sqrt{3}\rho/2)\phi_n(-\rho/2)\phi_j(\rho).
\end{eqnarray*}
In actual calculations, this formula can be further simplified. By defining $a_{n}=\sum_{m} c_{mn}\phi_m(0)$, Eq.(\ref{fermi_eigen_equation1}) can be simplified as
\begin{equation}
	\begin{split}
		-\frac{2\sqrt{\pi}}{g}a_{n}=\sum_{j}a_{j}\Big(&\frac{\sqrt{\pi}\Gamma(-\nu_n)}{\Gamma(-\nu_n+1/2)}\delta_{j,n}\\
		&-\int d\rho \Phi_n(\sqrt{3}\rho/2)\phi_n(-\rho/2)\phi_j(\rho)\Big),
	\end{split}
  \label{fermi_simplified_eigen_equation}
\end{equation}
where $v_{n}=(E/\omega-m-n-1)/2$ and $\Phi_{n}=\Gamma(-v_{n})e^{-\frac{x^2}{2l^2}}U(-v_{n},\frac{1}{2},\frac{x^2}{l^2})$, and $U(-v,1/2,x^2)$ is Kummer’s function.
Solving the large matrix equation (\ref{fermi_simplified_eigen_equation}), we can obtain both  $E$ and $\{a_{n}\}$. The coefficient $c_{mn}$ in the wavefunction $\Psi$ can be obtained via Eq.(\ref{fermi_eigen_equation1}). One can also prove that the anti-symmetry of $\Psi$ under the exchange $x_2\leftrightarrow x_3$ can be automatically guaranteed by the anti-symmetry of $f$-function in Eq.(\ref{f_symmetry}). 

In practically solving Eq.(\ref{fermi_simplified_eigen_equation}),   we have taken the cutoff as large as  $n_{\rm max} =60$, which allows the convergence of ground state energy up to the sixth digit (in unit of $\omega$).

\subsection{Four-body ($N=3$)}

Similarly, for $(1+3)$ system one can define the relative coordinates as 
\begin{equation}
	\begin{split}
r=x_2-x_1, \rho=\frac{2}{\sqrt{3}}[(x_3-(x_1+x_2)/2)],\\ \eta=\sqrt{3/2}[x_4-(x_1+x_2+x_3)/3].
    \end{split}
\end{equation} 
The exchange of $x_2$ with $x_3$ and $x_4$ gives another two sets of relative coordinates $\{r_+,\rho_+,\eta_+\}$ and $\{r_-,\rho_-,\eta_-\}$.

In the CoM frame, the four-body wave function can be expanded in terms of the harmonic eigen-states:
\begin{equation}
	\Psi(r,\rho,\eta)=\sum_{m,n,l} b_{mnl}\phi_m(r)\phi_n(\rho)\phi_l(\eta)
\end{equation}

Introducing an auxiliary function $f(r,\rho,\eta)\equiv U\Psi(r,\rho,\eta)$, and ensuring its exchange symmetry 
\begin{equation}
f(r,\rho,\eta)=-f(r_+,\rho_+,\eta_+)=-f(r_-,\rho_-,\eta_-), \label{f_symmetry2}
\end{equation} 
we can write $f$-function as  
\begin{equation}
	\begin{aligned}
		f=g\sum_{ijk}b_{ijk}\phi_i(0)&\big[\delta(r)\phi_j(\rho)\phi_k(\eta)-\delta(r_{+})\phi_j(\rho_+)\phi_{k}(\eta_+)\\		&-\delta(r_{-})\phi_j(\rho_-)\phi_{k}(\eta_-)
		\big].
	\end{aligned}\label{f-s}
\end{equation}
Note that Eq.(\ref{f_symmetry2}) guarantees the anti-symmetry of $\Psi$ under the exchange of fermion coordinates $\{x_2,x_3,x_4\}$.

Recalling the Lippmann-Schwinger equation (\ref{LS}) and defining $a_{nl}=\sum_{m} b_{mnl}\phi_m(0)$, we get
\begin{equation}
	\begin{split}
	-\frac{2\sqrt{\pi}}{g}a_{nl}=&\frac{\sqrt{\pi}\Gamma(-v_{nl})}{\Gamma(-v_{nl}+1/2)}a_{nl}\\&-\sum_{jk}a_{jk}\left(B^{(1)}_{nl,j}\delta_{k,l}+B^{(2)}_{nl,jk}\right),
	\end{split}
	\label{4body-s}
\end{equation}
in which
\begin{eqnarray}
	B^{(1)}_{nl,j}&=&\int dx \Phi_{nl}(\frac{\sqrt{3}x}{2})\phi_n(\frac{-x}{2})\phi_j(x),\\
	B^{(2)}_{nl,jk}&=&\iint dxdy \Phi_{nl}(\frac{x+2\sqrt{2}y}{2\sqrt{3}})\phi_n(\frac{5x-2\sqrt{2}y}{6}) \nonumber\\
	&&\ \ \ \ \phi_l(\frac{-\sqrt{2}x-y}{3})\phi_j(x)\phi_k(y),
\end{eqnarray}
where $\Phi_{nl}=\Gamma(-v_{nl})e^{-\frac{x^2}{2l^2}}U(-v_{nl},\frac{1}{2},\frac{x^2}{l^2}),v_{nl}=(E/\omega-n-l-3/2)/2$. Solving Eq.(\ref{4body-s}), we can obtain the energy $E$ and all $\{a_{nl}\}$. Further, the coefficients $\{b_{mnl}\}$ can also be obtained via Eq.(\ref{LS}).

In computing the integrals above, we have used the following identity:\begin{equation}
	\Gamma(-\nu_{nl})U(-\nu_{nl},1/2,x^2)=\sum_{k=0}^\infty\frac{(-1)^kU(-k,1/2,x^2)}{k!(k-\nu_{nl})}.\label{sum_k}
\end{equation}
It has the advantage that the energy dependence therein is fully incorporated in the parameter $\nu_{nl}$, as appearing in the denominator of above equation. This does not directly affect the real-space integrals. In this way, we can just compute and store the integrals of different $k$ once for all, and then sum over $k$ for different energies (or $\nu_{nl}$) to 
accelerate  the computation. In our practical calculations, we have taken the cutoff $n_{\rm max}=l_{\rm max}=60$ and the sum in (\ref{sum_k}) up to $k=30$, which allows the convergence of ground-state energy up to the fourth digit (in unit of $\omega$).

\end{document}